# The Role of Oligomeric Gold-Thiolate Units in Single Molecule Junctions of Thiol-Anchored Molecules

Edmund Leary,[†, ⊥, ‡, ∫]* Linda A. Zotti,[§] Delia Miguel,[‖] Irene R. Márquez,[†] Lucía Palomino-Ruiz,[‖] Juan Manuel Cuerva,[‖] Gabino Rubio-Bollinger,[⊥, ‡] M. Teresa González,[†] Nicolás Agrait,[†, ⊥, ‡]

† Instituto Madrileño de Estudios Advanzados (IMDEA), Calle Faraday 9, Campus Universitario de Cantoblanco, 28049 Madrid, Spain.
⊥ Departamento de Física de la Materia Condensada, Instituto "Nicolás Cabrera" and Condensed Matter Physics Center (IFIMAC), Universidad Autónoma de Madrid, E-28049, Madrid, Spain
‡ Laboratorio de Bajas Temperaturas, Departamento de Física de la Materia Condensada Módulo C-III, Universidad Autónoma de Madrid, E-28049, Madrid, Spain.
§ Departamento de Física Teórica de la Materia Condensada, Universidad Autónoma de Madrid, Madrid, E-28049, Spain.
‖ Departamento de Química Orgánica, Universidad de Granada, C. U. Fuentanueva, Avda. Severo Ochoa s/n, E-18071, Granada, Spain.
∫ Present address: Department of Chemistry, Donnan and Robert Robinson Laboratories, University of Liverpool, Liverpool L69 7ZD, U.K.

* E.Leary@liverpool.ac.uk

**ABSTRACT:** Using the break junction (BJ) technique we show that 'Au(RS)$_2$' units play a significant role in thiol-terminated molecular junctions formed on gold. We have studied a range of thiol-terminated compounds, either with the sulfur atoms in direct conjugation with a phenyl core, or bonded to saturated methylene groups. For all molecules we observe at least two distinct groups of conductance plateaus. By a careful analysis of the length behavior of these plateaus, comparing the behavior across the different cores and with methyl sulfide anchor groups, we demonstrate that the lower conductance groups correspond to the incorporation of Au(RS)$_2$ oligomeric units at the contacts. These structural motifs are found on the surface of gold nanoparticles but they have not before been shown to exist in molecular-break junctions. The results, while exemplifying the complex nature of thiol chemistry on gold, moreover clarify the conductance of 1,4-benzenedithiol on gold. We show that true Au-S-Ph-S-Au junctions have a relatively narrow conductance distribution, centered at a conductance of log(G/G$_0$) = -1.7 (± 0.4).

## Introduction

The gold-sulfur interface is one of the most important and well-studied of its kind in molecular electronics.[1] Thiols (as thiolates) are a mainstay for the stabilization and functionalization of gold nanoparticles and linking molecules to extended surfaces forming self-assembled monolayers (SAMs).[2] In particular, they have been one of the principle binding groups used to make contact to individual molecules via break junction (BJ) methodologies.[3][4] Despite this ubiquity, the nature of the interface often remains unclear, especially in BJ experiments, where atomistic analysis is generally impossible. There is a long-standing debate in Molecular Electronics regarding the origin and interpretation of the conductance distribution in thiol-anchored junctions using gold electrodes. Multiple conductance groups have been found in many experiments,[5][6][7] but their origins often remain controversial and hard to pin down precisely. A prime example is 1,4-benzenedithiol (BDT) which has been studied at the single molecule level for three decades, from the first molecular BJ experiment[3] in 1997 to the first single molecule thermopower study in 2007.[8] It is surprising, therefore, that even now there is a lack of consensus in the literature regarding its conductance in Au|molecule|Au single molecule junctions (SMJs). Its conductance has been quoted as ranging from close to 1 G$_0$ right down to values as small as 10$^{-4}$ G$_0$.[9][10][11][12] In other studies, BDT even fails to show a clear conductance histogram peak.[13][14]

In this paper we set out to investigate these apparent inconsistencies by primarily collecting large data sets returning several thousand molecular junctions per compound, separating out plateaus into various conductance groups, and also comparing thiol anchors with methyl sulfide. We

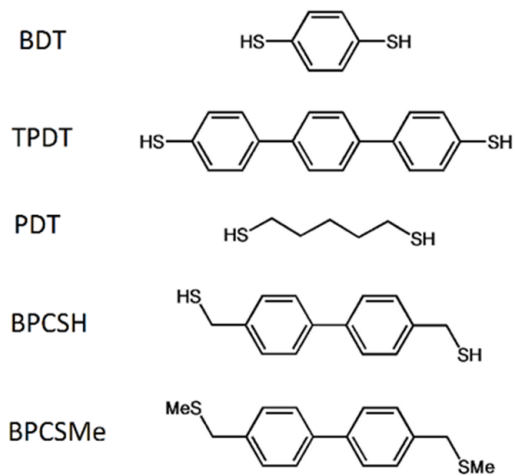

BDT

TPDT

PDT

BPCSH

BPCSMe

*Figure 1. Chemical structure of the compounds investigated. The distance (to one decimal place) between two gold atoms attached to each sulfur for each molecule is: 1.0 nm (BDT), 1.8 nm (TPDT), 1.3 nm (PDT), 1.7 nm (BPCSH and BPCSMe).*

have studied a range of simple thiol-terminated compounds (Figure 1) to look for any clear systematic differences, investigating both aromatic-thiols, where the sulfur is directly attached to a benzene ring (1,4-benzenedithiol; BDT and 4,4''-terphenyldithiol; TPDT) and compounds with saturated thiols, in which the sulfur is bonded to a methylene group, and whose reactivity may differ due to the lack of π-conjugation with the S atoms. We have also studied two compounds with the same backbone, one with thiols and the other with methyl sulfide (SMe) groups. This allows us to compare the behavior of molecules with identical backbones, but very different chemical properties at the anchor.

Our main conclusion is that for all thiol compounds studied, we find compelling evidence for the formation of short 'AuS' chains at the molecule-electrode interface. This suggests that molecules are associated on the surface in a manner akin to the surface of gold nanoparticles, which contain $Au(SR)_2$ structural motifs.[1]

**Experimental Methods**

We employed the break-junction method, which we have described previously,[15] using a home-built scanning tunneling microscope (STM).[16][17] Details of sample preparation and general methodology can be found in the supporting information section 2. Briefly, the STM tip is repeatedly driven in and out of contact with the surface and the current monitored as a function of the distance travelled. At values close to 1 $G_o$, the quantum of conductance (77.5 µS), small plateaus in G followed by a sharp drop indicate the final breakdown of the metallic junction. Below this, longer plateaus are observed in a percentage of junctions, which are indicative of molecular junction formation (Figure 2). If no molecule is wired, then an exponentially decaying 'tunneling-only' signal is observed (see Figures

S11-13 for examples). To process the data, we run an algorithm that filters molecular junctions from the tunneling-only junctions, and from this we create 2D density plots of the molecular conductance at all electrode separations. We further divide the plateau-containing traces based on the absolute conductance value of the plateau to separate traces with plateaus arising from different conductance groups as observed in the 2D histograms (see section 2.3 in the SI for details). This allows us also to correlate the occurrence of plateaus in different groups with each other. To determine plateau lengths, we measure the interval between two points in a single trace, starting at 0.3 $G_o$, and measuring until the last point at a given value well below the plateau. To calibrate the inter-electrode separation, we add 4 Å to the measured value for all junctions. Each junction should ideally be calibrated independently, but as this is impossible, the amount we add is effectively the minimum retraction determined from traces with no plateaus (see Plateau Calibration section, page S15). Doing so means we are unlikely to overestimate the break-down distances significantly post-calibration.

**Break Junction Results**

In the following sections, we will present our main experimental and theoretical findings which explain why $Au(SR)_2$ units (as depicted in Figure 6) are key to understanding the low-conductance plateaus in thiol molecular junctions. Figures 2 and 3 show examples of individual plateau-containing traces and the ensuing 2D histograms for each compound. From a quick inspection it is clear that each thiol-terminated compound displays more than one conductance group. The majority display two groups, whereas PDT clearly displays three (for further discussion of previous studies investigating multiple conductance groups, please see section 2.7 in the SI). BDT also shows a third group, although this only becomes clear after combining the results from two sample runs and separating the traces into the various groups (Figure S7). BPCSMe, with methyl sulfide anchor groups, displays just one group, and this was reproduced on more than three separate experimental runs (see SI section 2.8 for further details). The highest group (group 1) for each compound increases in length in correspondence with the molecular backbone (see Table 1 for a comparison of plateau group lengths as well as the calculated maximum junction lengths). Group 2 plateaus, on the other hand, generally only appear after a group 1 event. We have carried out a correlation analysis in order to quantify this behavior (see SI section 2.3 for further details). This analysis showed that for all compounds, approximately 50 % of all traces with a plateau displayed only a group 1 plateau. For BPCSH and TPDT, group 2 plateaus are practically exclusively found only after a group 1 event. For BDT and PDT, more group 2/3-only traces were found, but the general behavior is similar.

The results of the plateau separation/correlation analysis on BPCSH are shown in Figure 4, where we compare these with the results of BPCSMe. We have carried out the same type of analysis for all other compounds (see sections 2.6-

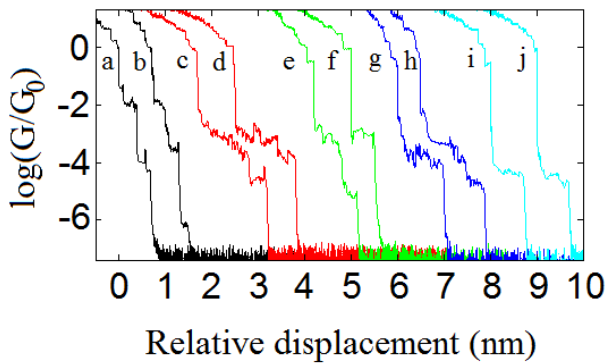

*Figure 2. Examples of individual traces with plateaus for each compound. a, b: BDT; c, d: TPDT; e, f: PDT; g, h: BPCSH; i, j: BPCSMe.*

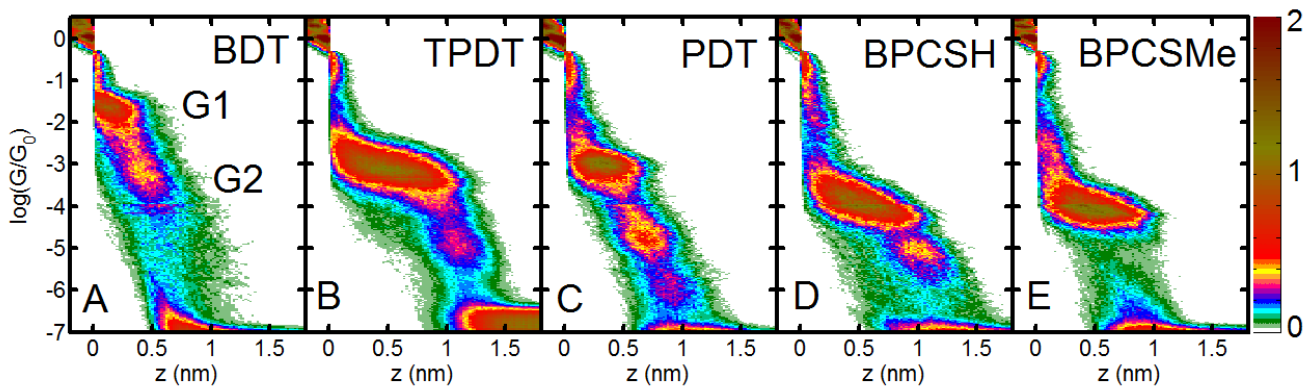

*Figure 3. 2D histograms of all plateau-containing traces for each compound, and only the background tunneling traces have been removed. The number of traces in each histogram and percentage of the total is; BDT: 2059 (28 %), TPDT: 4724 (50 %), PDT: 2780 (27 %), BPCSH: 1886 (20 %), BPCSMe: 2492 (37 %).*

2.8 in the SI for conductance/distance data, and a detailed breakdown of the percentages of plateaus within each group). It is noticeable that group 1 of BPCSH strongly resembles the one group found for BPCSMe. There is, however, a reproducible difference in the plateau lengths, with those of the thiol being consistently longer by approximately 0.2 nm. This can be understood simply due to the weaker nature of the Au-SMe bond[18][19] compared to Au-S, which produces no significant deformation of the contacts. The calibrated plateau length ($L_p$) data for all compounds for group 1 and group 2 are given in Table 1. The main values were obtained by fitting a single Gaussian to the length distribution histograms, and extracting the maximum of the peak. The values in parentheses are the 95th percentiles. Comparison with the calculated junction lengths for each reveals that the 95th percentiles for group 1 agree well with the calculated length, although this is also true of the group 2 peak maxima. The 95th percentiles of group 2, however, generally extend well beyond the calculated distance, and in the case of BDT this is by 0.5 nm. For BDT/PDT, group 3 extends to 0.7 nm beyond this value. If group 1 corresponds to a fully-stretched junction, group 2/3 would be hard to reconcile with the typical behavior of gold atomic contacts, which at room temperature are expected to deform by 2-3 Å at most before rupture occurs.[20]

For each compound, the ratio of group 1 to group 2 conductance is very similar, about 1.5 orders of magnitude (or about a factor 30). This is also the case between group 2 and group 3 for PDT and BDT. As far as we are aware, such low conductance plateaus for these compounds have not been reported previously. The difference in junction length between groups 1 and 2 for each compound (obtained by subtracting the maximum in the Gaussian fit for group 1-only traces from the maximum in the fit of group 2-containing traces) is also quite constant at between 0.2-0.3 nm. For PDT the difference between group 2 and group 3 is also 0.2 nm. This behavior points towards an atomic difference between groups 1, 2 and 3 for all compounds, and also strongly suggests a common origin.

It is important to highlight that the conductance of group 1 of BPCSH ($\log G/G_o = -4.0$) is higher than BPCSMe (-4.1), whereas group 2 for BPCSMe (-5.1) is evidently much lower. We have previously compared OPE3 compounds (where 3 represents the number of aryl rings) with thiol and SMe anchor groups. We found that the SMe had a lower conductance than the thiol (-3.8 for the thiol, -4.6 for SMe).[21] We have also recently compared a family of diphenyl porphyrins with the same anchor groups, and find the same trend in conductance, this time measuring -4.4 for the thiol and -4.8 for the SMe.[22] These results match the trend we observe for group 1 of BPCSH and the SMe analogue. Based on this, it is highly unlikely, therefore, that group 2 of BPCSH corresponds to junctions with a single BPCSH molecule wired S to S between the electrodes (i.e.

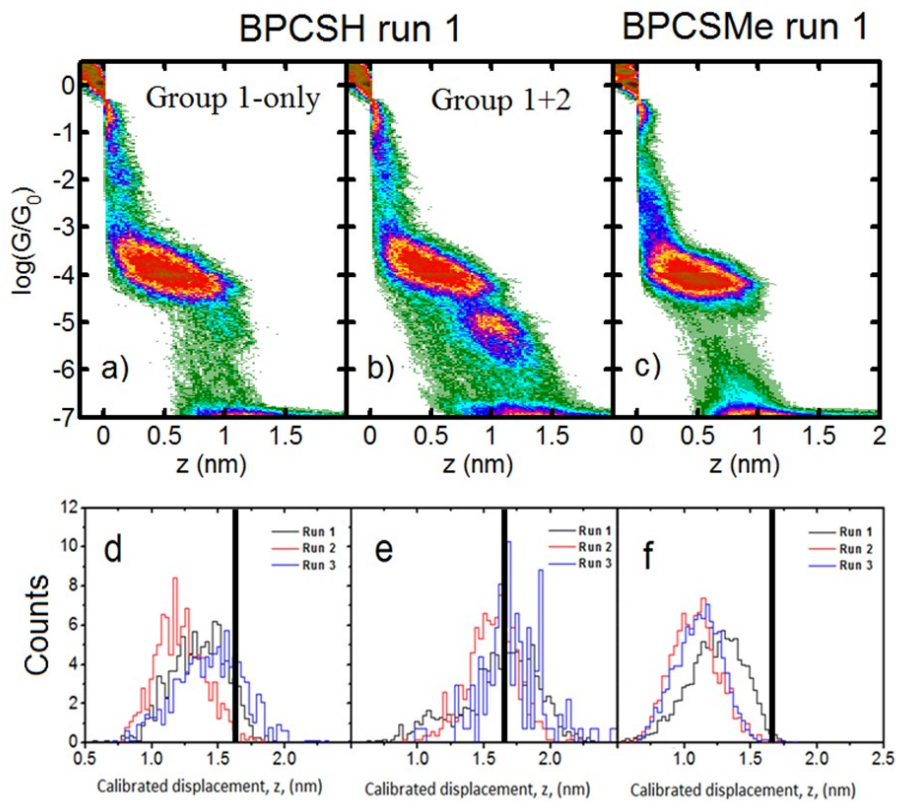

*Figure 4. a) Separated traces containing only group 1 plateaus for run 1 of BPCSH. b) Separated traces containing a group 1 and group 2 plateau for run 1 of BPCSH. c) All plateaus recorded during run 1 of BPCSMe. N.b, z-values in a-c are without calibration. d) Junction length distributions of group 1 for run 1 for each run of BPCSH. e) Junction length distributions measuring until the end of group 2 for each run of BPCSH. f) Junction length distribution for BPCSMe. The black vertical lines represent the expected single molecule junction length.*

it cannot be a typical SMJ). It is hard to imagine a situation in which the thiol group becomes so strongly decoupled from the electrodes whilst still maintaining a full Au-S bond leading to a distinct conductance group. We thus assign group 1 of BPCSH to that of a single molecule in an Au-S-CH$_2$(Ph)$_2$CH$_2$-S-Au junction. It is, however, straightforward to understand the conductance distribution observed in group 1 of all compounds (which is about one order of magnitude) as indicative of the various conformations and binding sites that the molecule can adopt during stretching.

We noticed that, unlike for BPCSH, the peak for BPCSMe could be better fitted by the sum of two gaussians separated by approximately a factor two (see Figures S15-16), and it is possible to interpret this as being due to one and two-molecule junctions respectively (wired in parallel, but non-interacting). If so, then the lower of the two peaks would be that of the SMJ, which is located at log(G/G$_0$) = -4.2. The width of the group 1 peak for BPCSH is generally larger than for BPCSMe, with the mean FWHM over the runs being 0.94 and 0.68 respectively (or 0.42 for the lower peak in the two Gaussian fit). If group 1 for BPCSH arrises predominantly from SMJs, then this result implies thiols have only a slightly larger conductance distribution over their methyl sulfide counterparts.

For BDT and TPDT, the conductance values of group 1 and group 2 decrease with a beta-value of 0.40 Å$^{-1}$ (1.7 per phenyl). This agrees well with the analogous amine-terminated series measured by Venkataraman et al.[23] The conductance of group 1, particularly with TPDT, falls in line with our previous measurements on the longer OPE3 dithiol, which shows a slightly lower conductance at log(G/G$_0$) = -3.8.[17][24]

## Discussion

To elucidate the origin of group 1 and group 2, we can firstly rule out a transition between the gold surface and the layer of ambient adsorbates present in the experiment, which we have observed for larger C$_{60}$ anchor groups. Although this could produce a similar G(z) profile, it would most likely affect both SH and SMe groups to some extent (and others such as RNH$_2$, which is not observed[25]). The distinct absence of low conductance plateaus for BPCSMe therefore rules this out. We can also side-line the idea of multi-molecular junctions associated through interactions such as π-stacking.[26] PDT clearly cannot π-stack, and neither would we expect BDT to have sufficiently strong enough interactions to behave in this way. The low percentage of molecular junctions formed also goes against this, and we actually consistently find higher percentages of plateaus (junctions) for the SMe compound, indicating that multiple molecule effects are more likely for this compound (this is actually implied by the histograms fitting better to the sum of two gaussians as shown in Figures S15-16).

the Fermi energy) of BPCSMe and BPCSH is $1.22 \times 10^{-4}$ $G_0$ and $1.04 \times 10^{-4}$ $G_0$ respectively. We can compare these values

| Molecule | $G_{G1}$ (log(G/G_0)) | $G_{G2}$ (log(G/G_0)) | $L_{p(G1)}$ (nm)[b] | $L_{p(G2)}$ (nm) | Calculated Length (nm)[c] |
|---|---|---|---|---|---|
| BDT | -1.71 | -3.52 | 0.84 (1.09) | 1.01 (1.53) | 0.99 |
| TPDT | -3.15 | -4.85 | 1.48 (1.81) | 1.66 (2.01) | 1.81 |
| PDT | -3.04 | -4.68 $G_3 = -5.97$ | 0.99 (1.22) | 1.33 (1.64) $G_3 = 1.55$ (1.87) | 1.27 |
| BPCSH | -4.00[a] | -5.1[a] | 1.26[a] (1.69)[a] | 1.56[a] (1.94)[a] | 1.73 |
| BPCSMe | -4.20[a] | | 1.05[a] (1.40)[a] | | 1.73 |

Table 1. Summary of the main experimental results for each molecule and each group. Values in parentheses are the 95th percentiles. [a]Mean value over all runs. [b]$L_p$ values are the maximum values of a Gaussian fit to the plateau length distributions, and include 0.4 nm calibration distance. [c]See section 2.4 in the SI for the junction lengths including additional Au-S units.

Finally, although dimerization through disulphide bond formation may occur, it is not a valid explanation for groups 2 and 3 in general. The conductance ratio between group 1 and 2 should change from compound to compound, which it does not, and secondly, dimerization would not produce a constant 0.2-0.3 nm length difference between successive groups.

**Theoretical Section**

In order to obtain insight, we have carried out detailed density functional theory (DFT) calculations combined with non-equilibrium Green's function (NEGF) technique and the HOMO-LUMO gap correction using a procedure previously described[27][28] (see Supporting Information section 3 for further details). We focussed primarily on the BPCSH/SMe pair as their comparison provides the best opportunity to elucidate the mechanisms at work. We tested different plausible binding scenarios that may occur during the stretching of BPCSH junctions. In Figure 5 we show the geometries analyzed for BPCSH (a-d) and for BPCSMe (e). For an enlarged view of the different contacts please see Figure S21. In order to make a like-for-like comparison we first compared both compounds directly in a geometry they can both adopt. We found that the SMe group prefers binding to undercoordinated atoms, in line with previous calculations,[29] and hence chose to analyse junctions in which the terminal S atoms are bound to the last gold atom of a pyramidal electrode in a top orientation (geometry (a) for BPCSH and geometry (e) for BPCSMe). The zero-bias transmission curves are shown in Figure 5 (lower panel, BPCSH - solid black line BPCSMe - cyan dashed line) where, despite the energetic positions of the HOMO and LUMO resonances for BPSH being about 1 eV more positive compared to BPCSMe, the two curves cross each other very close to the Fermi level. This means we expect the measured conductance to be similar for both at full stretch. The theoretical conductance (given by the transmission at to the experimental log/linear-histogram maxima ($1 \times 10^{-4}$

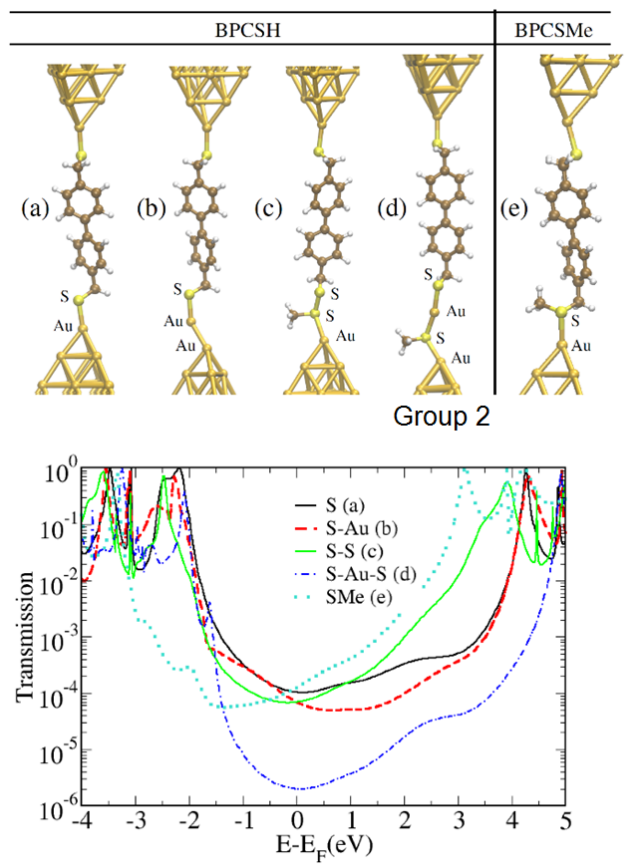

Figure 5. Upper panel: Optimized geometries of BPCSH in which the S atom at one end is bound to the apex atom of a $Au_{26}$ pyramid (a), to an Au adatom (b), to an $SCH_3$ group (c), to an Au-$SCH_3$ group (d) and BPCSMe in a top-top binding position (e). In (c) and (d) the $CH_3$ group replaces a full biphenyl molecule.

Lower panel: transmission as a function of energy for the geometries (a)-(e) of the upper panel.

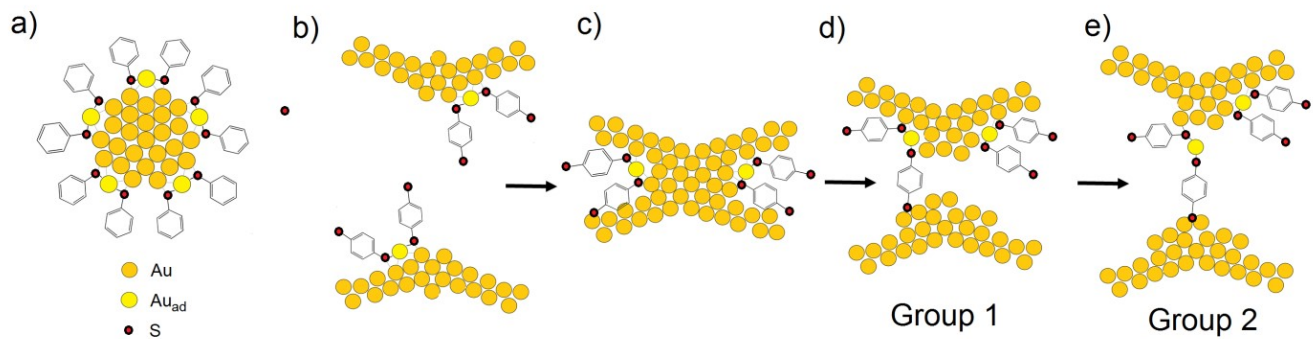

<image_description>Figure showing panels a) through e) with gold nanoparticle and molecular junction representations. Panel d) labeled "Group 1" and panel e) labeled "Group 2".</image_description>

Legend:
- ● Au
- ● Au$_{ad}$
- ● S

d) Group 1
e) Group 2

*Figure 6. a) Representation of a thiol-capped gold nanoparticle showing the RS-Au(I)-SR surface motif. b-e) Proposed model of thiol behaviour inside a gold-electrode BJ experiment using BDT as an example.*

$G_0$ / 5 x 10$^{-5}$ for BPCSH and 6 x 10$^{-5}$ $G_0$ / 5 x 10$^{-5}$ for BPCSMe respectively). Within the limits of both experiment and theory we can conclude, therefore, that on a like-for-like binding basis, in a geometry most likely to occur at full elongation, the conductance of BPCSH is very similar to BPCSMe. Our calculations also support the conclusions drawn from the experimental junction length distribution analysis. They strongly suggest thus that group 2 of BPCSH is not related to a simple electrode-molecule-electrode binding scenario. This also rules out the previously proposed hollow-to-top (or high-to-low Au-S coordination transitions) as being responsible,[30][7] which requires junctions associated with top-coordination to be longer than those associated to hollow-coordination.

Even though we do not expect significant Au-aryl interactions in our measurements, as seen elsewhere,[31][32] we decided to try and force the situation in our calculations just to see what might happen. We therefore explored the situation for a compressed form of both junctions by reducing the inter-electrode separation, forcing the molecules to sit higher up one of the electrodes, placing the central biphenyl group partially in contact with the gold via its π-system (Figure S18). We found that in both cases an increase in transmission upon compression occurred. If we take the results of both molecules in isolation, the sliding model may actually seem like a reasonable explanation of high to low conductance state transitions. This would hold for BPCSH (for which we observe high and low conductance groups experimentally) but not, we stress, for BPCSMe, where there are no additional conductance groups between the main group and the noise at 10$^{-7}$ $G_0$. Assigning the group 2 of BPCSH to its full-stretched geometry implies the conductance of the fully-stretched form of BPCSMe to be < 10$^{-7}$ $G_0$ (our experimental noise level). This would contradict the theory presented so far, and would also be strongly at odds with previous comparison between -SH and -SMe terminated compounds performed with the same technique.[21]

To understand what is going on, we note that the measured difference in elongation between high and low states is on the order of 2-3 Å. This means we should focus on configurations in which an extra one to two atoms are inserted at one of the contacts. We considered a geometry in which one thiol S atom is kept in the same top position, while the other is connected to an extra Au atom protruding from the electrode apex (Figure 5b), simulating the extrusion of a gold adatom. We then considered binding to a sulfur atom from a second neighboring molecule in the form of a disulfide linkage. To replicate the second molecule, we used an S-CH$_3$ group rather than a whole molecule to reduce computational time (Figure 5c). For the validity of this approximation, please see section 3.5 in the SI. Looking at the transmission curves in Figure 5 for these junctions, we find no significant difference in transmission at the Fermi level between the standard top-top geometry of (a) and those of (b) and (c) which could account for the observed difference, and we side-line these geometries as possible origins of the lower group 2 state. We suggest instead that they could contribute to the overall distribution of group 1.

In a report from Jadzinsky et al. the total structural determination of a large gold-thiolate cluster using x-ray crystallography revealed a particularly special arrangement of Au and S atoms in which atoms were found lifted from the bulk forming a bridge between several thiolate ligands, forming RS-Au(I)-SR units.[33] This arrangement has been corroborated through evidence from low temperature STM, in particular on Au(111) surfaces covered with methanethiol in which pairs of molecules have been imaged bridged by single gold adatoms.[34] It was proposed by Strange et al. that such motifs could explain the large conductance variation in thiol-bound junctions[35] and so we turned our attention to this configuration. To reduce computation time, we used a single –SMe group in place of the second molecule. In the lower panel of Figure 5, we show the transmission curve for this configuration as the blue dashed line. This junction clearly produces a large drop in transmission at the Fermi level, giving a conductance of 2 x 10$^{-6}$ $G_0$, which is almost two orders of magnitude lower than the top-top configuration of (a). This fits well with the behaviour seen experimentally for group 2, for which the linear histogram maximum lies at 3 x 10$^{-6}$ $G_0$. Our theoretical Au-Au distance of 2.0 nm also matches the experimental group 2 95$^{th}$ percentile value, also 2.0 nm. Overall, therefore, we find that the insertion of AuS units between the main electrode and the wired molecule reproduces the experimental observations very well for BPCSH, unlike the other configurations anaylzed.

Moreover, additional 'AuS' units provide a succinct explanation for the origin of the additional groups for the other thiols measured. For PDT/BDT, where we observed three groups, multiple 'AuS' units may form either on both sides of the junction, or as chains only on one side. Our results suggest that this effect is universal for all thiol compounds which are, at least, not sterically prevented from forming such Au(I)-bridged dimers. Other factors may play a role in determining the formation of oligomeric AuS units, such as surface mobility. Our results further highlight that in the absence of oligomerization, thiols would display a only a slightly larger conductance distribution compared to other anchor groups,[36] as evidenced by the distribution of plateaus within group 1.

We have also carried out calculations on BDT (Figure S19-20) as a way of cross checking our theory with the already published data.[35] In spite of using different theoretical implementations, we find the ratio of our calculated transmission values for the single BDT junction and that with the extra 'AuS' unit is very close to that published by Strange et al. (Figure S20). Specifically, we calculate an Au-BDT-Au junction conductance of $6.28 \times 10^{-3}$ $G_o$, and an Au-S-Au-BDT-Au junction conductance of $1.57 \times 10^{-4}$ $G_o$. The experimental values (at $V_{bias}$ = 0.2 V) are $2 \times 10^{-2}$ $G_o$ (group 1 log max), $1 \times 10^{-2}$ $G_o$ (linear max), and (for group 2) $3 \times 10^{-4}$ $G_o$ (log max) $1 \times 10^{-4}$ $G_o$ (linear max). We have sketched schematically a typical situation we expect during a break junction experiment for BDT in Figure 6.

For highly mobile molecules, such as short alkanes, it is likely that self-assembly into $Au(SR)_2$ units takes place rapidly, most likely on both sides of the junction, during elongation, making chain formation at either side of the contact possible. Indeed, as mentioned, STM studies of methanethiol reveal it to form these structures on gold under low surface coverages.[34] For larger molecules, we imagine this is a more difficult process due to reduced mobility, which is perhaps hindered further by rough surfaces as generated in BJs. For BPCSH and TPDT, however, the conductance of junction with multiple AuS units would fall below our noise level, and further studies using larger amplification would be needed to test this.

### Experiments with propanethiol

We have further investigated what happens when just a monothiolated compound (1-propanethiol, PT) is present on the gold. Figure 7b shows the 2D histogram for a gold sample exposed to PT. We do not find long plateaus, as for the other dithiol compounds, but there are similar features in the region between $\log(G/G_o)$ = -1 to -3, which result from short (1-2 Å) plateaus. In particular, we observe a clear protuberance centred around $\log(G/G_o)$ = -2.4. This behavior is markedly different to unmodified gold (Figure 7a), which tends to display much fewer plateaus in this region, and only a narrow feature located at $\log(G/G_o)$ = -0.8. We also can see on the BPCSH sample (Figure 7c) a diffuse region of plateaus which extends to approximately $\log(G/G_o)$

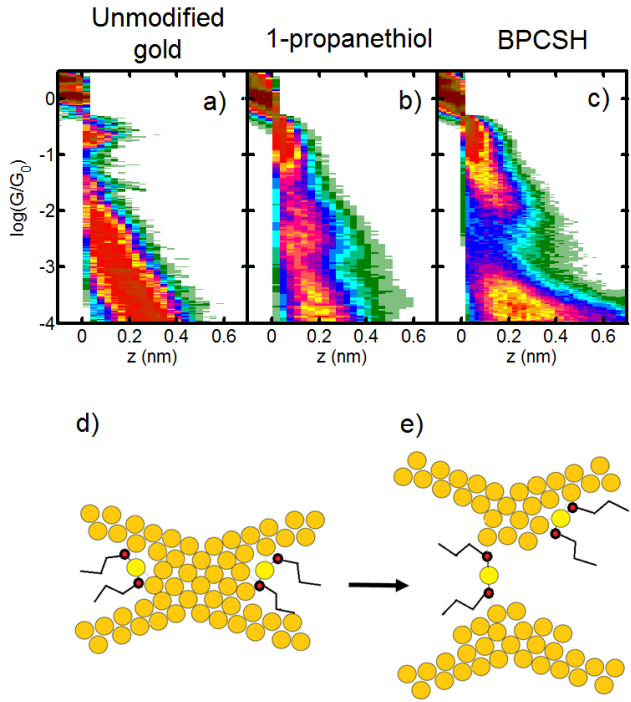

*Figure 7. (top panel) High conductance region of traces a) without thiols present, b) in the presence of 1-propanethiol and c) BPCSH. (lower panel) d and e: model of the type of junction which may occur in the presence of any thiol group.*

= -2, resembling the PT sample. This is also the case for the alkanedithiols measured previously.[37] The calibrated length of the plateaus for PT is between 6-8 Å, meaning the junctions in this region fit with the picture proposed in Figure 6d and e in which Au(I)-bridged dimers link across the junction in the manner shown. These observations further validate our interpretation of group 2/3 by providing direct evidence that small structural entities are present within the junction when thiols are adsorbed which could become incorporated between the electrodes and the vertically-wired dithiol compound.

Furthermore, we decided to check what happens if PT is added to a solution of BPCSMe and subsequently exposed to gold. The results (Figure S17) show clearly that only one main conductance group is present, identical to that found without addition of the monothiol. This supports our hypothesis that to observe 'AuS' chain formation after group 1 requires both the vertically-wired molecule and the second participating molecule to be thiolate in nature.

### Conclusion

In summary, we have demonstrated that oligomeric 'AuS' units produce the low conductance groups of plateaus during the stretching of thiol-anchored molecular junctions. The frequency of these plateaus in relation to the overall percentage of junctions implies that thiol molecules are associated on the surface similar to how they are on the surface of gold nanoparticles, where RS–Au(I)–SR structural motifs are commonly found. The RS–Au(I)–SR units appear stable and robust to stretching. These motifs, well-known on the surface of thiol-capped gold nanoparticles, have until now not been recognised in 'single' molecule junctions. Our break junction experiments on the thiol-anchored molecules alone strongly suggest the formation of the oligomeric Au-S units at the molecule-electrode interface, however further confirmation of this is demonstrated by the comparison of the two biphenyl compounds, one with thiol anchors, and the other with methylsulfides. The SMe compound robustly displayed just one conductance group, which appears very similar to group 1 of the thiol. The plateau length and conductance of both molecules strongly suggest they are due to fully-stretched single molecule junctions, which implies that groups 2 and 3 originate from a situation unique to thiols. Coadsorption of 1-propanethiol (PT) with the SMe compound did not produce a second group of plateaus. The presence of PT, however, does significantly modify the region between $\log(G/G_0)$ = -1 to -3, producing short plateaus which are consistent with the presence of RS-Au(I)-SR dimer junctions. Similar plateaus are generally seen for each dithiol.

Overall, our results highlight the complex nature of structures present in thiol-anchored molecular junctions. By applying the RS–Au(I)–SR over-layer model, however, our results allow us to pin down an unambiguous conductance signature for true single molecule junctions of 1,4-benzenedithiol, which we determine have a conductance equal to $\log(G/G_0)$ = -1.7 (± 0.4).

## ASSOCIATED CONTENT

Materials and methods, details of distance calibration, control experiments, computational details and additional information on theoretical calculations. The Supporting Information is available free of charge via the Internet at http://pubs.acs.org.

## AUTHOR INFORMATION

**Corresponding Authors**

E.Leary@liverpool.ac.uk
linda.zotti@uam.es

**Author Contributions**

The manuscript was written through contributions of all authors. All authors have given approval to the final version of the manuscript.

## ACKNOWLEDGEMENTS

This work was supported by MICINN (Spain) (MAT2014-57915-R; MAT2017-88693-R; Nanofrontmag-CM S2013/MAT-2850) and by the EC FP7 ITN "MOLESCO" Project Number 606728. EL acknowledges IMDEA. MTG acknowledges MICINN for her Ramón y Cajal contract. LAZ was funded by the Spanish MINECO through the grant MAT2014-58982-JIN. I.R.M. thanks UGR for a postdoctoral scholarship. IMDEA Nanociencia acknowledges support from the 'Severo Ochoa' Programme for Centres of Excellence in R&D (MINECO, Grant SEV-2016-0686).

We acknowledge useful discussion with Herre SJ van der Zant and Juan Carlos Cuevas.

## Notes
The authors declare no competing financial interests.